# Multiple Bent Crystal Reflections for Efficient Beam Collimation in Frontier Colliders

M.Yu.Chesnokov, Yu.A.Chesnokov\*, V.A.Maisheev,
Yu.E.Sandomirskiy, A.A.Yanovich, I.A.Yazynin

*Institute for High-Energy Physics (IHEP) in National Research Center "Kurchatov Institute", Protvino, Moscow region, 142281 Russia*

\**e-mail: chesnokov@ihep.ru*

Abstract.

The Large Hadron Collider (LHC) uses a multi-stage collimator system to absorb the growing halo of circulating beams to protect and ensure reliable operation of superconducting magnets. A similar system is planned for the Future Circular Collider (FCC). In anticipation of the LHC operation with high luminosity, research is being conducted to improve the collimation system. Studies have shown that one of the solutions to improve beam collimation is to use channeling in a short curved crystal, which acts as a primary collimator, throwing particles deep into the secondary collimator by channeling. This system is very sensitive to the angular position of the crystal and possible vibrations of different nature. In this paper, we propose a different approach to crystal collimation based on the volume reflection of particles from curved crystallographic planes in a sequence of crystals. The positive qualities of this scheme are substantiated and a multi-strip crystal device capable of implementing it is proposed.

The Large Hadron Collider (LHC) uses a multi-stage collimator system to absorb the growing halo of circulating beams to protect and ensure reliable operation of superconducting magnets [1]. A similar system is planned for the Future Circular Collider (FCC) [2]. Primary LHC collimators made of carbon fiber composites deflect halo particles as a result of Coulomb scattering, thereby increasing their impact parameters to secondary collimators. The interaction of protons with the collimator material leads to the formation of diffraction protons, which can fly out of the walls of the collimators and get lost in the magnets, which limits the efficiency of cleaning the existing collimation system. In anticipation of the LHC operation with high luminosity, research is being conducted to improve the collimation system. Studies [3] show that one of the solutions to improve beam collimation is to use channeling in a short curved crystal, which acts as a primary collimator, throwing particles deep into the secondary collimator due to channeling (Fig.1).

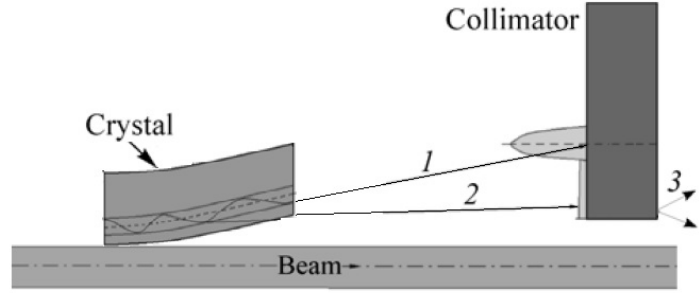

Fig.1 Scheme of beam collimation using a short crystal, 1-peak of channeled particles that effectively deflected, 2-fraction of dechanneled particles, 3-losses on the collimator.

Since the critical channeling angle (in silicon single crystals of (110) orientation) is quite small at high energies, $\theta_c$ = 2.5 μrad for a 6.5 TeV beam in the LHC, and $\theta_c$ = 0.92 μrad for a 50 TeV beam in the FCC, the system is very sensitive to the angular position of the crystal and possible vibrations of different origin. One of the types of vibration can be a phenomenon known since the 80s [4], which is caused by the recoil force when bunches of particles interact with a crystal in the channeling mode. Due to the high energy of each particle, the particle deflected by channeling transmits a noticeable transverse momentum to the crystal. According to estimates [5], even $10^6$ particles in a bunch can create appreciable oscillation amplitudes, which will impair the efficiency of crystal collimation.

In this regard, another approach to crystal collimation based on the volume reflection of particles from curved crystallographic planes may be useful. This phenomenon was discovered recently in experiments [6-7], and it expands the boundaries of the use of curved crystals on accelerators. The volume reflection was first predicted in [8] in computer simulation by the Monte Carlo method and later described in detail in analytical form in [9]. Volume reflection is caused by the interaction of an incoming relativistic particle with the potential of a curved atomic lattice and occurs at a short length in the region tangent to the curved atomic plane, leading to a deflection of the particle in the direction opposite to the bend. The probability of reflection $eff$ is high and for positive particles of TeV energies is close to unity. In Ref. [10] it is shown that the efficiency of the reflection process is limited by the value of an alternative process called volume capture, the probability of which is equal to:

$$P_{vc}(R) = \frac{1.39 A U_0^{1/4} J_p}{2^{7/4}\sqrt{\pi} E_0^{1/4} \varepsilon_{max} d^{1/2} X_0^{1/2}} \left(\frac{R}{R_c} - \frac{\kappa_1}{\kappa_c}\right) \simeq 1 - eff \qquad (1)$$

where $R$ is the bending radius, $E_0$ is the particle energy, $U_0$ is the planar potential barrier, $\mathcal{E}_{max}$ is the maximal value of planar electric field, $d$ is the interplanar distance, $X_0$ is the radiation length, $R_c$ is the critical channeling radius and $A = 11\ MeV$, $J_p = 1.49$, $\kappa_c = 0.186$, $\kappa_1 = 0.13$ are constants for silicon single crystal. For determination of $U_0$, $\mathcal{E}_{max}$, $R_c$ values we used the potential function from x-ray measurements [9]. In particular, $R_c = 10.83$ m and 83.33 m for $E_0 = 6.5$ and 50 TeV, correspondingly. For these particle energies equation is valid for $R/R_c$ less than 40-50.

The reflection parameters of a particle with an energy of 6.5 TeV and 50 TeV, the average angle of reflection $\alpha$, the **rms** value, and the efficiency calculated from the model [9,10] with an account of Eq. (1) are shown in Fig. 2ab. It should be noted that the calculations performed according to the model are in a good agreement with experiments at particles energies up to 400 GeV (see detail comparison in [10]).

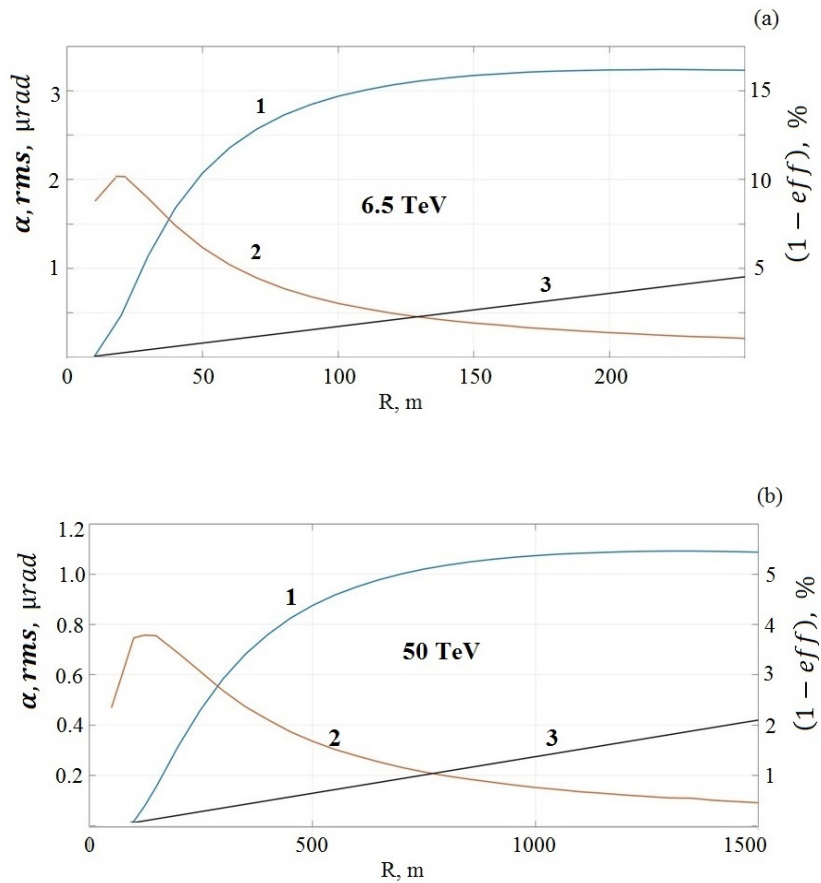

Fig.2 ab  The dependence of the reflection parameters versus the bending radius of the crystal: curve 1 – the average angle of reflection $\alpha$, 2- the **rms** of angular distribution, 3 - the value (1-$eff$).

The reflection angle in a single crystal, as can be seen from the figure, is small, about 1.4 $\theta_c$, but can be amplified in a sequence of crystals (Fig.3). In the framework of UA9 activity at

CERN, various variants of multicrystal systems were studied [11-14] to multiply the deflection of particles. One of the successful options for ultra-high energies can be a device developed in the IHEP [15], tested on a beam of 400 GeV protons [14].

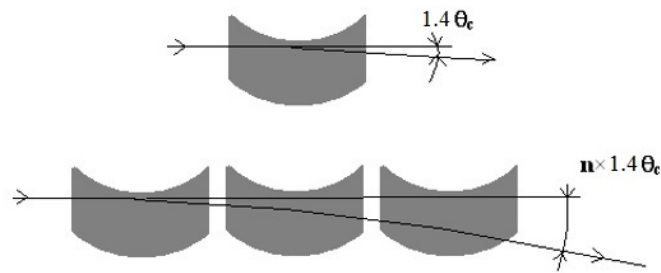

Fig.3 Amplification of the deflection angle of particles during successive reflection on a chain of well aligned crystals.

The schematic diagram of the crystal deflector and its photo are shown in Fig.4ab. The deflector was made of a silicon plate measuring 70 ×15 ×5 mm. the Large faces of the crystal plate were parallel to the planes of the crystal (111), while the end face was perpendicular to the axis <110>. In contrast to the method based on the use of an external force generated by the holder, the proposed method uses internal stresses created by mechanically applied grooves on the surface of a thick crystal plate. The depth of the triangular grooves in our case was about 1.1 mm. The Bending of individual strips of 2 mm length formed between the grooves was produced by the deformation of the surface layers due to the Twyman effect [16]. Due to the thick common base of the crystal deflector, the relative position of the strips on the surface, both angular and spatial, is much better than when using an external bending device. Fig. 4c shows an effective (over 90%) deflection of the proton beam by this crystal according to the results [14].

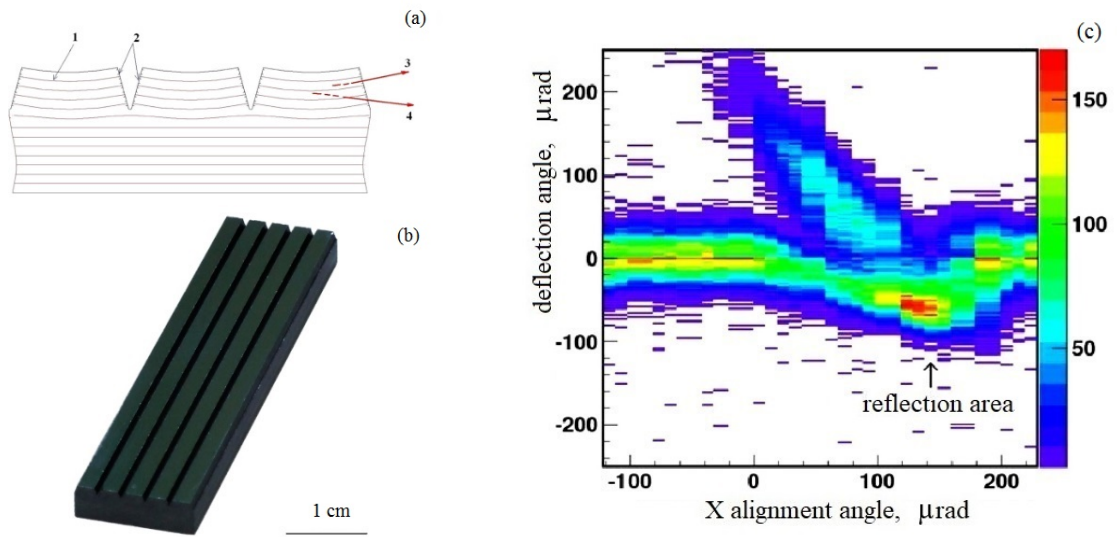

Fig.4 a-Schematic representation of a curved multi-strip crystal formed by periodic grooves on the surface of a thick crystal. (1) curved crystallographic planes, (2) rough surfaces of grooves, (3) a particle deflected due to channeling, and (4) a particle reflected multiple times by curved planes. b) a photo of a silicon crystal plate with periodic grooves. c) - effective deflection of protons 400 GeV due to multiple reflection in curved strips, according to [14].

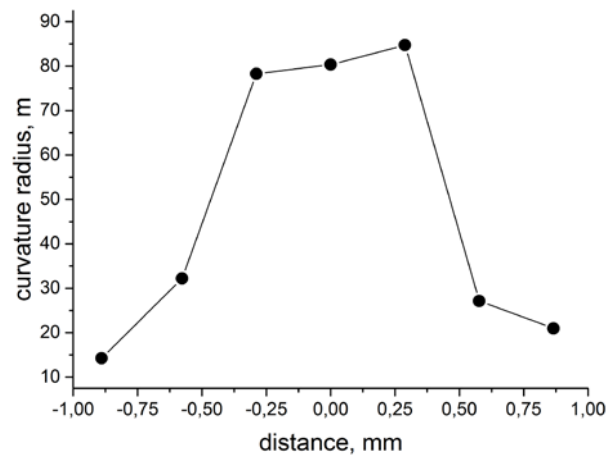

Fig.5. Bending radius of the crystal strip along its length, according to measurements [17].

A parallel x-ray beam was used to study the bending of individual strips and their mutual alignment at the Kurchatov synchrotron radiation source KISI [17]. Research has shown that this design, a series of curved strips formed between large grooves on a thick plate, is so well mutually aligned that it is suitable for collimating the LHC proton beam and even a planned 50 TeV FCC accelerator using multiple particle reflection. Fig.5 shows the bending radius of each silicon strip along the length of the crystal. In the center of the strip where the reflection occurs, the radius is constant, about 100 m, which is optimal for the energy of the LHC. There is no problem optimizing the device for 50 TeV energy by varying the depth of the grooves and the length of the crystal strips.

Using the SCREPER program [18], the Monte Carlo method was used to calculate the deflection of particles with energies of 6.5 TeV and 50 TeV in a system of several crystal strips. For an energy of 6.5 TeV, a crystal with five strips, each 3 mm long, with a bending radius of 100 m and a bending angle of 30 μrad was selected. For an energy of 50 TeV, the crystal has five strips of 5 mm long, with a bending radius of 800 m and a bending angle of 6 μrad. The results of calculations for 6.5 TeV energy are shown in Fig.6a. It can be seen that in a wide range of angles (about 30 μrad), the crystal effectively deflects the beam due to multiple reflection. The multiple reflection efficiency on multiple curved strips is high, about 92 % for 6.5 TeV LHC energy and 95 % for 50 TeV FCC energy. The corresponding deflection angle for multiple reflections on five strips is 15 μrad and 5.2 μrad, and it can be increased due to the axial orientation of the device, as demonstrated in [14]. Fig.6b shows the effect of calculating the axial deflection at reflection for an energy of 6.5 TeV. With the axial orientation of the crystal, the average scattering angle and the angular width of the beam increase several times, compared to the planar case. This property is very important for removing radiation loads on the walls of secondary collimators. In this case, the crystal not only improves the efficiency of collimation, but also protects expensive secondary collimators from destruction.

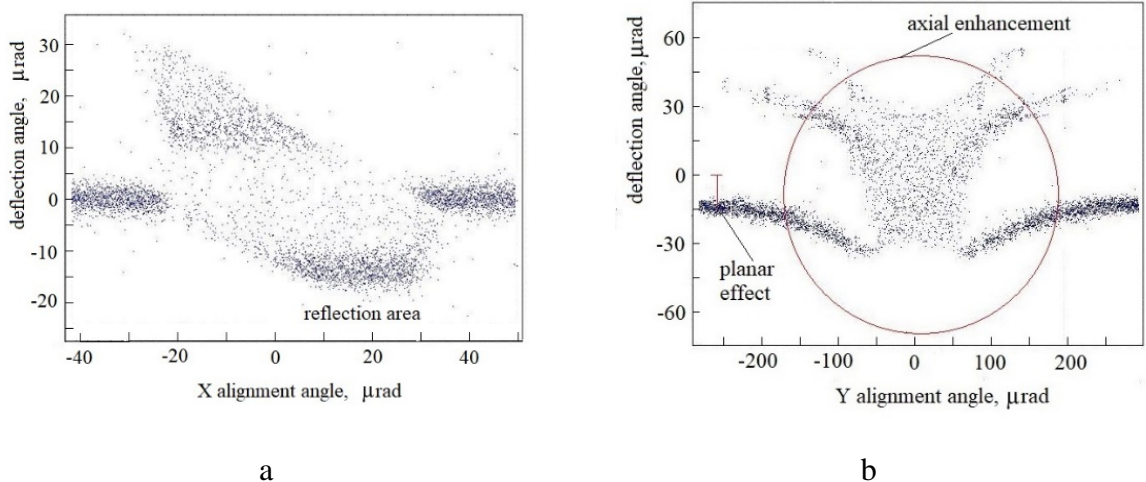

a                                                                                   b

Fig.6 a-deflection of particles with an energy of 6.5 TeV by a multi-strip device in the mode of planar reflection of particles. b-amplification of the reflection angle in the axial orientation of the crystal device.

Note the additional positive qualities of using multiple reflection, in comparison with the use of channeling:

Wide working range at the angles – within the entire angle of bending of the crystals, this is about 30 μrad for the LHC and 6 μrad for the FCC. This means that there is no strong sensitivity to vibration, as in channeling. There is also less requirement for a goniometric device. There is no need to adjust the crystal angle for each store cycle. It is enough to adjust the angle once and then move the crystal only linearly, as usual collimators. There is less requirement for the perfection of the crystal, since the reflection occurs in the center of the crystal at a small length of about 1.2 $R \times \theta_c$, a fraction of a millimeter (Fig.7), while when channeling, the particles make dozens of oscillations along the entire length of the curved crystal. Therefore, the radiation resistance will be higher and the long-term stability will be higher as a result. These arguments suggest the promise of this method of beam collimation at ultra-high energies.

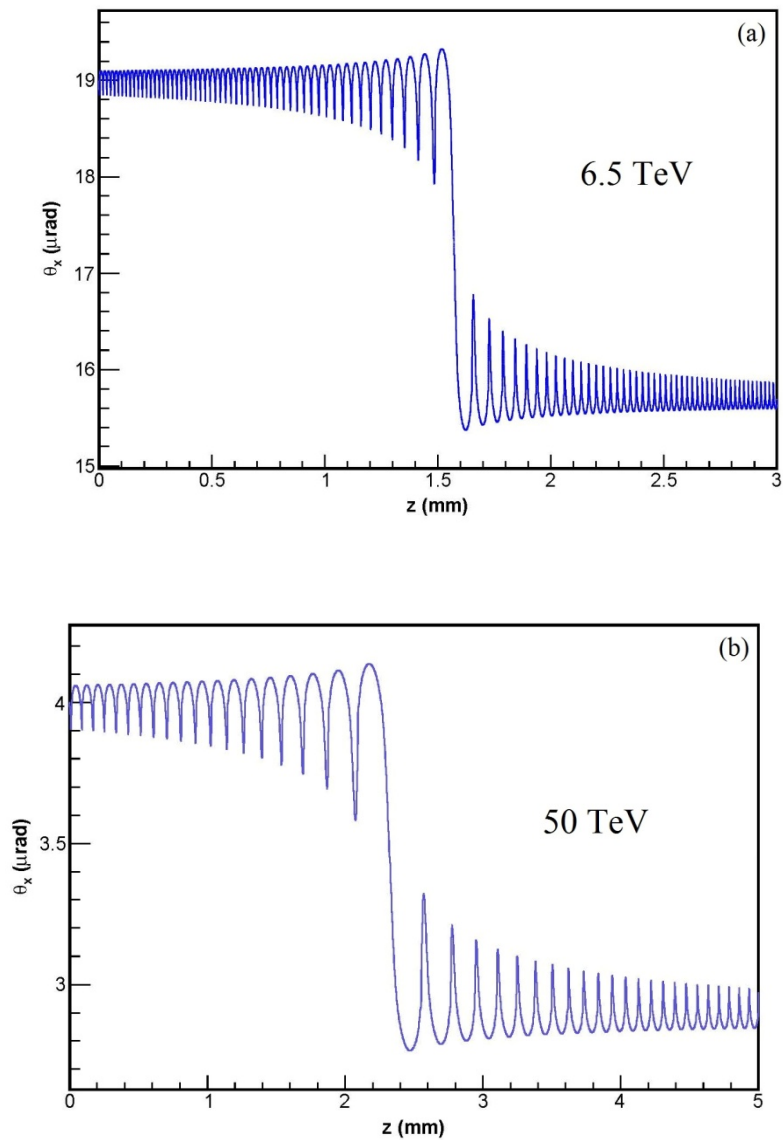

Fig.7 The behavior of the angle $\theta_x$ of particle in a short curved crystal during volume reflection.

It should be noted that the mode of beam deflection due to multiple reflection in mulstrip crystal devices was used for beam collimation in accelerators at lower energies, at the U70 accelerator at an energy of 70 GeV in the IHEP [19,20] and at an energy of 1 TeV at the Tevatron accelerator in Fermilab [21]. These studies have shown that even at lower energies, collimation on reflection is no worse than when channeling in short crystals.

The work was supported by the Russian Science Foundation, grant 17-12-01532.